\documentclass{acm_proc_article-sp}

\usepackage{amsfonts}
\usepackage{amsmath}
\usepackage{graphicx}
\usepackage{algorithmic}
\usepackage{algorithm}
\usepackage{multirow}
\usepackage{url}

\usepackage[usenames]{color}
\definecolor{emphcolour}{rgb}{0.8,0.2,0.2}
\newcommand{\PF}[1]{}	

\usepackage[usenames]{color}
\definecolor{emphcolour2}{rgb}{0.2,0.2,0.8}
\newcommand{\SL}[1]{}	

\begin{document}
%
\title{Generic Multiplicative Methods for Implementing Machine Learning Algorithms on MapReduce}

\author{
\numberofauthors{3} 
\alignauthor 
Song Liu\thanks{This work was done when Song Liu was with University
of Bristol}\\
       \affaddr{Sugiyama Laboratory}\\
       \affaddr{Tokyo Institute of Technology}\\
       \email{Song@sg.cs.titech.ac.jp}
\alignauthor
Peter Flach\\
       \affaddr{Intelligent Systems Laboratory}\\
       \affaddr{University of Bristol, UK}\\
       \email{Peter.Flach@bristol.ac.uk}
\alignauthor
Nello Cristianini\\
       \affaddr{Intelligent Systems Laboratory}\\
       \affaddr{University of Bristol, UK}\\
       \email{Nello.Cristianini@bristol.ac.uk}
}

\maketitle

\begin{abstract}
In this paper we introduce a generic model for multiplicative algorithms which is suitable for the MapReduce parallel programming paradigm. We implement three typical machine learning algorithms to demonstrate how similarity comparison, gradient descent, power method and other classic learning techniques fit this model well. Two versions of large-scale matrix multiplication are discussed in this paper, and different methods are developed for both cases with regard to their unique computational characteristics and problem settings. In contrast to earlier research, we focus on fundamental linear algebra techniques that establish a generic approach for a range of algorithms, rather than specific ways of scaling up algorithms one at a time. Experiments show promising results when evaluated on both speedup and accuracy. Compared with a standard implementation with computational complexity $O(m^3)$ in the worst case, the large-scale matrix multiplication experiments prove our design is considerably more efficient and maintains a good speedup as the number of cores increases. Algorithm-specific experiments also produce encouraging results on runtime performance. 

\category{D.1}{Programming Technique}{General}
\category{F.1}{Analysis of Algorithms and Problem Complexity}{General}

\terms{Algorithms}

{\bf Keywords:} Multiplicative Model, Machine Learning, MapReduce
\end{abstract}

\section{Introduction}
\label{sec:intro}
In order to deal with increasing dataset sizes, Machine Learning algorithms are required to be implemented in very large scale. However, upscaling learning algorithms is not always straightforward. In recent years, the MapReduce paradigm\footnote{US Patent Class: 712/203} \cite{Dean04mapreduce:simplified} and its open-source implementation Hadoop \cite{White10} has drawn increasing attention from industry for its remarkable capability of processing large-scale data and straightforward functional programming representation. Questions have been raised that whether MapReduce paradigm can scale up learning algorithms in a succinct fashion that a wide range of learning algorithms can benefit from. \SL{address the problem that we are going to solve}

Among the major algorithms, similarity-based classification approaches occupy a dominant position in the Machine Learning research field. Earlier attempts have made a variety of algorithms in this kind available for the MapReduce paradigm. Solutions to several key models including $k$-Means and K-Nearest Neighbor (KNN) have been proposed at an early stage and good accelerating performances were reported \cite{zhao09,stupar10}. Individual implementations are also documented for Support Vector Machines on Graphics Processing Units (GPUs) \cite{Catanzaro08} and Locality sensitive hashing (LSH) for Google News Personalization \cite{das2007}. Because of this broad interest, two similarity and distance-based learning algorithms are selected as our upscaling target. \SL{why similarity-based}

Efforts have also been made for establishing a generic model that solves several algorithms at the same time. Gillick et al.\ grouped several machine learning algorithms into three categories: Single Pass, Iterative, and Query-based \cite{gillick06}. Recent work has been done by Chu et al.\ \cite{Chu06} where 10 different learning algorithms have been introduced. Inspired by their study, it is realized that a generic model of upscaling several Machine Learning algorithms may be found, and the implementation work can be greatly reduced, thus, discovering a generic model of several learning algorithms is our main interests in this paper. \SL{why generic model}

Liu et al.\ proposed a methodology for web-scale Non-Negative Matrix Factorization in \cite{liu10} using a multiplicative approach as described by Lee and Seung \cite{Lee01algorithmsfor}. An interest has been brought that two multiplication models can be generalized from their implementation for scaling up other learning algorithms, and it is also an inspiration for us to adopt multiplicative methods on our targeting learning problems. \SL{why multiplicative}

\hyphenation{Page-Rank}\hyphenation{Page-Ranks}
The primary goal, therefore, of this research is to upscale a range of machine learning algorithms including Non-Negative Matrix Factorization (NMF), Support Vector Machines (SVM) and PageRank by utilizing a generic multiplicative model on  MapReduce paradigm. 
\PF{You need a bit more here: Why are we doing this? Why is this new? What has been done before? \\Also, hasn't Google already upscaled PageRank to MapReduce? }

The paper is organized as follows. Related works are discussed in the next section, and we then introduce the problem settings and solutions of these three algorithms, after which a theoretical study will be adopted for extracting multiplicative models from three algorithms which defines our core problem in the next section. The methodologies employed for extending these algorithms in large scale will be illustrated in two parts: \textit{a)} common computation components parallelization and \textit{b)} algorithm-specific settings optimization. Results from a wide range of experiments follow, and a brief conclusion is summarized in the final section.

\section{Related Work}
In recent years, implementing Machine Learning algorithms on MapReduce Paradigm have been widely discussed in literatures. Having been proposed in 2004, MapReduce is believed to be the large-scale parallel data processing engine in Google for a wide range of services (e.g. webpage indexing and page repository hosting) \cite{Dean04mapreduce:simplified}. \cite{das2007} reported that the News Service in Google also takes suggestions from learning algorithms running on MapReduce. However, there is no direct evidence published by Google showing how PageRank \cite{Brin98theanatomy} is operating with MapReduce. \SL{MapReduce and Google}

Efforts by individuals or groups other than Google have also contributed a number of ideas for running Machine Learning algorithms on MapReduce. One of the most recent efforts by Liu et al.\ indicates a novel way of upscaling Non-Negative Matrix Factorization on MapReduce by using the multiplicative method where the iterative update approach described in \cite{Lee01algorithmsfor} is adopted as a series of matrix multiplication. Several multiplication strategies are developed at different stages. In order to balance the load of servers and maximize the parallelization, a partitioning strategy is performed for large matrices. However, partitioning a huge matrix into single rows/columns and combining into multiplicative permutations may consume considerible computational resources. But considering the problem setting they need to handle (extremely sparse matrices), it is acceptable in most cases. In contrast, a more generalized plan is illustrated in section \ref{sec:method}. \SL{Major Reference for Multiplicative Methodologies}

Rather than upgrade one single individual algorithm at one time, generic frameworks for Machine Learning algorithms are also discussed in \cite{gillick06,Chu06}. Gillick et al.\ investigated a taxonomy of standard machine learning algorithms, and data processing patterns were taken into primary consideration which leads to three major groups of algorithms: Single Pass, Iterative Learning and Query-based Learning, revealing both advantages and limitations of the three learning paradigms. It is realised immediately after this research that a large set of algorithms can be phrased as MapReduce fashion by following the same pattern. Their work also illustrates the benefit that simplified MapReduce program representations offer to the Machine Learning community. 

The same year, a larger collection of algorithms have been implemented on MapReduce by means of the Statistical Query Model \cite{kerns1998} in \cite{Chu06}. Statistical query learning uses statistical properties of the data rather than individual examples to perform noise-tolerant learning. Given such theory and an objective function, learning algorithms are typically optimization algorithms that can be written in a summation form which naturally fits the MapReduce paradigm. Algorithms first calculate the sufficient statistics and gradient from a statistical query oracle and then aggregate them over all data points, thus datasets can be distributed among cores and the Map function is responsible for examining partial gradients while the Reduce stage checks through all Map-generated data for aggregation. 
While this method forms the foundation of the iterative update implementation in our study, it does not use statistical queries but generally borrows the idea of ``Summation Form" to calculate the gradients directly from individual examples. \SL{Foundation theory for our study}

SVM implementation, as a notable exception, has been illustrated and adapted efficiently on MapReduce for Graphics Processors, but failed to migrate to PC clusters since generally the traditional Sequential Minimal Optimization (SMO) \cite{platt98} method may require more than ten thousand iterations on a medium sized dataset \cite{Catanzaro08}. This iterative process causes significant start-up overhead for general Hadoop PC clusters. Chu et al.\ also proposed their own SVM implementation under summation form, however, they fail to explain how to handle a gigantic kernel matrix for large-scale dataset. In this paper, we believe, Quadratic Programming, the naive form of SVM can be constructed by using two models extracted from the previous research reported in \cite{Chu06} and \cite{liu10}. \SL{Why we are interested in SVM on PC MapReduce}

\section{Adapted Algorithms and Solutions}
In this section, definitions of three learning problems with their typical solutions are given. 

\subsection{Non-Negative Matrix Factorization}
The definition of NMF is as follows:
\newtheorem{theorem}{Definition}
\begin{theorem}
\label{def:nmf}
\PF{You need to consistently use mathbf for matrices here} \SL{fixed}
Given $\mathbf{A}\in{\mathbb{R+}^{m\times n}}$ and a positive integer $k\leq min(m,n)$, find a factorization of $\mathbf{A}$ into $\mathbf{W}\in{\mathbb{R+}^{m \times k}}$ and $\mathbf{H}\in{\mathbb{R+}^{k \times n}}$, such that divergence function $\mathcal{D}\mathbf{(A||\tilde{A}})$ is minimized, where $\mathbf{\tilde{A}=WH}$ is the reconstructed matrix from the factorization,
and the divergence function is defined as 
\begin{displaymath}\mathcal{D}(\mathbf{A||\tilde{A}})=\sum_{i,j} (\mathbf{A}_{i,j} - \tilde{\mathbf{A}}_{i,j})^2=||\mathbf{A-WH}||^2\end{displaymath}
\PF{I suspect there's a square missing in the above formula.} \SL{fixed}
\end{theorem}
From the probabilistic view, NMF methods can be divided into different types in which each element $\mathbf{A}_{i,j}$ is an observation from the distribution whose mean is $\mathbf{\tilde{A}}_{i,j}$. In this paper, we only consider one popular NMF, Gaussian NMF (GNMF) by taking 
\begin{displaymath} \mathbf{A}_{i,j} \sim \mathit{Gaussian}(\mathbf{\tilde{A}}_{i,j, },\sigma ^{2}) \end{displaymath}
\PF{Shouldn't that be the other way around: $\tilde{A}$ consists of noisy samples from $A$?}
\SL{This is the direct definition borrowed from Lee and Seung}

GNMF is solved by Lee and Seung \cite{Lee01algorithmsfor}, using a multiplicative approach:
\PF{explain what you mean by a multiplicative approach; also explain $.*$ below (I suspect that stands for Matlab pointwise matrix product, but that isn't really standard mathematical notation)}
\SL{look better?}
\begin{equation}
\label{eq:nmflinear}
\mathbf{H}_{i,j} \leftarrow \mathbf{H}_{i,j} \frac{(\mathbf{W^{T}} \mathbf{A})_{i,j}} {(\mathbf{W^{T} W H})_{i,j} } ;
\mathbf{W} \leftarrow \mathbf{W}_{i,j} \frac{(\mathbf{A H^{T}})_{i,j}} {(\mathbf{W H H^{T}})_{i,j} }
\end{equation}
under which the divergence $\mathbf{||A-WH||}^2$ is non-increasing after each update. 

\subsection{Support Vector Machines}
The one-norm soft-margin SVM with fixed bias can be defined as:
\begin{theorem}
\label{def:svm}
\begin{displaymath}\mathrm{maximize\ }  W(\mathbf{\alpha})= \sum_{i=1}^{l}\alpha_i -  \frac{1}{2}\sum_{i,j=1}^{l} y_{i}y_{j}\alpha_{i}\alpha_{j} K(\mathbf{x_i,x_j})\end{displaymath}
\begin{displaymath}\mathrm{subject\ to\ }  0\leq\alpha_i \leq C  \mathrm{\ and\ }  i=1,2,\ldots, l \end{displaymath}
\end{theorem}
This definition uses the fixed bias so that the constraint $\sum_{i=1}^{l} \alpha_i y_i =0$ does not need to be explicitly included (for a formal definition please refer to \cite{nello00}). 

According to the definition \ref{def:svm}, it is equivalent to minimize: 
\begin{displaymath}-2 \sum_{i=1}^{l}\alpha_i + \sum_{i,j=1}^{l} y_{i}y_{j}\alpha_{i}\alpha_{j} K(\mathbf{x_i,x_j})\end{displaymath}
we set the partial derivatives wrt.\ $\alpha$ to 0, so that 
\begin{displaymath}-2y_i + 2\sum _{j=1} ^{l} \alpha_j \mathbf{K(x_i,x_j)} =0 \end{displaymath}
\PF{I don't think there should be a factor 2 here, since you have 1/2 in the original formula}
or
\SL{Hmmm...that?}
\begin{displaymath}\mathbf{G \alpha}=y\end{displaymath}
where $\mathbf{G_{i,j}=K(x_{i},x_j)}$. The optimal solution of parameter $\alpha^*$ is then given by 
\begin{equation}
\label{eq:linearsvm1}
\mathbf{\alpha}^*=\mathbf{G}^{-1}y
\end{equation}
and can be obtained by iterative gradient descent approach which will be introduced in section \ref{sec:grad and power}. \PF{say where} \SL{here}

\subsection{PageRank}
PageRank \cite{Brin98theanatomy} follows a recursive definition as follows:
\begin{theorem}
\begin{displaymath}PR(p_i)= \frac{1-d}{N}+d \sum_{p_j\in{M(p_i)}} \frac{PR(pj)}{L(p_j)} \end{displaymath}
\end{theorem}
where $p_1, p_2, \ldots, p_n$ are PageRanks for webpages, $d$ is a damping factor between 0 and 1 which simulates how quickly a ``Random Surfer" is getting tired during surfing, $N$ is the total number of webpages, while $L(p_j)$ is the total number of outlinks  for a single webpage. 

PageRank \textit{de facto} represents the eigenvector for a stochastic matrix in a Markov chain with its maximal eigenvalue, 1. This problem can be solved by a very effective approach called ``Power Method".
For a transition probability matrix $\mathbf{P}$ of a directed graph $\mathbf{G}$, and $\mathbf{\pi}$ donates the stationary probabilities of Markov Chain, so $\mathbf{\pi}$ satisfies:
\begin{displaymath}\mathbf{\pi = P \pi}\end{displaymath} 
Moreover, the $\mathbf{\pi}$ is the principal eigenvector of matrix P, with its maximal eigenvalue 1. 
The stationary probabilities $\mathbf{\pi}$ can be obtained by power method \cite{Arasu02pagerankcomputation} which employs the iterative multiplication as follows:
\begin{equation}
\label{eq:pageranklinear}
\mathbf{\pi^{k+1} = P \pi^k }
 \end{equation}
\section{Multiplicative Models}
In order to parallelize these three algorithms by means of a generic approach, two types of common ``multiplicative components" are extracted from given solutions. 

\subsection{Similarity Comparison and Distance Computations}
Similarity comparisons (e.g. dot product and distance computations) is a general calculation involved in a variety of learning algorithms. Dot product for two matrices with rows in the form of vectors, can be proceeded using the following matrix multiplication and transposition:
\begin{displaymath}\mathbf{A \cdot B=AB^{T} }\end{displaymath}
According to  \eqref{eq:nmflinear}, each NMF update requires the multiplication of two large matrices for similarity comparison. Both $\mathbf{A H^{T}}$ and $\mathbf{H H^T}$ calculate the inner products of $n$-dimension vectors on ``column features" of matrix $\mathbf{A}$. Similarly, $\mathbf{W^T A}$ and $\mathbf{W^T W}$ gives the similarity measure of $m$-dimension vectors on the ``row features" of matrix $\mathbf{A}$. Similar story can also be found in \eqref{eq:linearsvm1} where kernel matrix can also be formed by multiplying two matrices with training vectors and its transposition. 

Euclidean distances can be handled in the same way as both Euclidean distance and dot product can be written in ``Summation Form".

This type of matrix multiplication is characterized by its large high-dimension of input matrices and high-density of output matrix which cause severely storage problems. In the next section, we will demonstrate how these problems can be avoided in our implementation by using high storage capacity and data locality feature of MapReduce. 

\subsection{Gradient Descent and Power Iterative Method}
\label{sec:grad and power}
For many optimization problems, the aim is to learn a parameter vector $\theta$ from a linear system generalized $\mathbf{A}$ and a sequence of observations $\mathbf{y}$. 
%
In general we have $\mathbf{y} = \mathbf{\theta^T A}$ and thus $\mathbf{\theta^*} = \mathbf{A^{-1} y}$. However, matrix inversion is computationally costly, and instead methods such as gradient descent are used. Addressed in \cite{Chu06}, these algorithms can be adapted to ``Summation Form" as well. 

As a typical Quadratic Programming (QP) problem, SVM can be represented in this form, and solved by a gradient descent method, in which each single parameter $\alpha_i$ can be updated by an increment:
\begin{displaymath}\delta \alpha_i^t = \eta \frac{\partial{\mathbf{W(\alpha ^t)} }}{\partial{\alpha^i}} =  \eta ( - y_i \sum _{j=1} ^{l} {y_j \alpha_j  \mathbf{K(x_i, x_j)} } +1) \end{displaymath}
Particularly, the gradient $\mathbf{G=\Delta W(\alpha)}$ of $\mathbf{W}$ with respect to $\mathbf{\alpha} $ can be expressed by linear algebra:
\begin{equation}
\label{eq:linearsvm}
\mathbf{G = \eta .* ( -y .* DK +1)}
\end{equation}

where $\mathbf{D}$ is a vector $\mathbf{D}=\{D_1, D_2, \ldots, D_l\}$, and $\mathbf{D_i}=y_i \alpha_i$.
Obviously, for multiplying a large dense matrix $\mathbf{K}$, the computational complexity of \eqref{eq:linearsvm} is dominated by the matrix multiplication $\mathbf{DK}$, which suggests us this component can greatly benefit from parallelized implementation. 

Similarly, Power Method demonstrated in \eqref{eq:pageranklinear} shows all the calculation in one iteration can be done by a simple multiplication. Common component extracted from these two algorithms can help us to handle these specific types of computations on large data.

In this category of matrix multiplication, two operands often varies in size. For instance, SVM gradient descent involves a large and dense kernel matrix multiplied with a much smaller column vector, so that the parallelism adopted in similarity comparison multiplication cannot be used in this case for efficiency consideration.

\section{Methodology}
\label{sec:method}
In this section, we demonstrate how these two types of matrix multiplication can be adapted on MapReduce paradigm.

\subsection{General Matrix Multiplication}
It has been proved that Partitioning is an efficient solution to large-scale Matrix Multiplication on MapReduce \cite{liu10}. We further generalise their approach by adopting the classical block matrix multiplication method. 

The typical method for distributed matrix multiplication is to use block matrix multiplication in which each operator matrix is partitioned across row or column, so that a large computation problem be divided and conquered.
\begin{theorem}
\label{def:generalmulpart}
For a matrix multiplication $\mathbf{C=A \times B}$, where $\mathbf{A} \in{\mathbb{R}^{a \times b}}$ and $\mathbf{B} \in{\mathbb{R}^{b \times c}}$, and $\mathbf{C} \in{\mathbb{R}^{a \times c}}$, a \textbf{Partition Schema}\footnote{Hadoop also uses the word ``Partition" to represent the idea of ``Shard" we discuss further down. However, it has no relationship with partition multiplication } $m,n,k$ can be introduced so that A and B can be partitioned as
\begin{displaymath}\mathbf{A} =
 \begin{pmatrix}
  \mathbf{A_{1,1}}  & \cdots & \mathbf{A_{1,n}} \\
  \vdots   & \ddots & \vdots  \\
  \mathbf{A_{m,1}} & \cdots & \mathbf{A_{m,n}}
 \end{pmatrix}
 \mathbf{B} =
 \begin{pmatrix}
  \mathbf{B_{1,1}}  & \cdots & \mathbf{B_{1,k}} \\
  \vdots   & \ddots & \vdots  \\
  \mathbf{B_{n,1}} & \cdots & \mathbf{B_{n,k}}
 \end{pmatrix}
\end{displaymath}
\noindent
Therefore, the result matrix $\mathbf{C}$ is also a block matrix with $m$ row partitions and $k$ column partitions, and each block $\mathbf{C_{\alpha,\beta}}$ can be obtained by 
\begin{displaymath} \mathbf{C_{\alpha,\beta}=\sum _{\gamma=1} ^{n} A_{\alpha \gamma} B_{\gamma,\beta}}\end{displaymath}
where $\alpha = 1\ldots m, \gamma = 1\ldots n, \beta = 1\ldots k$
\end{theorem}
\subsubsection{Partition-Summation Process}
\begin{figure}
\centering
\includegraphics[scale=0.35]{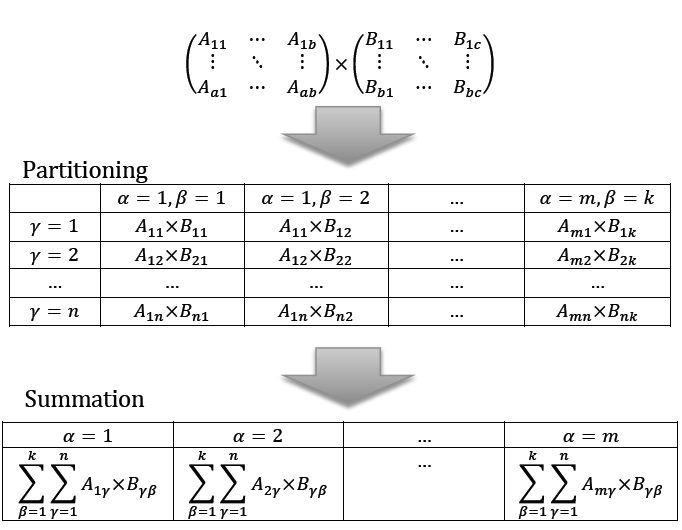}
\caption{General Matrix Multiplication with Partitioning and Summation process}
\label{fig:multi1}
\end{figure}

We summarize the algorithms described in definition \ref{def:generalmulpart} as a two-stage process illustrated in Figure~\ref{fig:multi1}. The first stage reads input $\mathbf{A}$, $\mathbf{B}$ and outputs partitioned sub-matrices grouped by their ``Partition Identifiers'' $\langle\alpha, \beta, \gamma, i\rangle$ while the second stage performs the actual multiplication of partitioned matrices and sums them up into the final result. Partition Identifier $\alpha, \beta$ shows which part of the summation result belongs to in result matrix while the $\gamma$ is used for identifying the sub-multiplication groups. In practice, this procedure is often split into two cascaded MapReduce jobs, in which the first job is responsible for partitioning and grouping matrices, and the second concentrates on multiplying two sub-matrices and summing up the intermediate results generated from the earlier stage. Algorithm details are shown in Algorithms \ref{alg1}, \ref{alg:part:red}, \ref{alg2} and \ref{alg3}. 

\begin{algorithm}
\caption{Partition Mapper (for matrix $\mathbf{A}$)}
\label{alg1}
\begin{algorithmic}
\REQUIRE row $\mathbf{a_i \in \{a_1, a_2, \ldots, a_b\}}$
\REQUIRE partition schema $\{m,n,k\}$
\STATE $\alpha \gets i / m$
\IF {$\alpha > m$}
\STATE $\alpha \gets m$
\ENDIF
\STATE $step \gets b/k$
\FOR{$i = 1$ \TO $b$ \textbf{step} $step$} 
\STATE $start \gets i; end \gets i+step$
\STATE $\gamma \gets i/step$

\IF {$\gamma > n$}
\STATE $gamma \gets n $
\STATE $end \gets b$
\ENDIF

\STATE $\mathbf{sub} \gets subvector(start,end,\mathbf{a_i})$

\FOR{$\beta = 1$ \TO $k$} 
\STATE $emit(\langle\alpha, \beta, \gamma, i\rangle, \mathbf{sub})$
\STATE $\beta \gets \beta + 1$
\ENDFOR

\STATE $i \gets i + 1$
\ENDFOR
\end{algorithmic}
\end{algorithm}

\begin{algorithm}
\caption{Partition Process (Reducer)}
\label{alg:part:red}
\begin{algorithmic}
\REQUIRE partition group identifier $\langle\alpha, \beta, \gamma, i \rangle $
\REQUIRE individual rows $\{\mathbf{A}_j\} \in \mathbf{A}_{sub}, \{\mathbf{B}_k\} \in \mathbf{B}_{sub}$
\STATE $emit(i, \{\mathbf{A}_{sub}, \mathbf{B}_{sub}\})$
\end{algorithmic}
\end{algorithm}

\subsubsection{Partition Schema}
\label{sec:partsch}
Partition schema often has a very significant impact on performance, for example, increasing $m$ and $k$ duplicates the partitioned matrices for more sub-multiplication groups, while increase $n$ may generate more intermediate results after partition multiplications are calculated. 
Generally, the computational complexity of this approach is $O(m \times n \times k)$, however, considering the sparsity of $\mathbf{A}$ and $\mathbf{B}$, the computational complexity is often lower in practice.  

\subsubsection{Sharding, Hashing and Computational Locality}

In MapReduce, communication cannot be made between nodes except the ``Shuffling" stage , at which step intermediate results generated from Map Stage are transferred to the nodes referring to the \textbf{Computational Locality} (i.e. the place where their final computation will be made). Each piece of intermediate results grouped by computational locality is called \textbf{Shard} in MapReduce.  

In this case, locality can be maximized when sub-matrices are multiplied where the summation operation will be proceeded, so that all computation can be done without data transfer after partitioning. When each shard is emitted from its original Mapper, a function $h$ determines where each piece of shard is going to be located. Using each Partition Identifier as its input, a naive form of $h$ can be written as:
\begin{displaymath} h_{naive}(\alpha,\beta,\gamma) = \alpha \mod p\end{displaymath}
where $\alpha = 0,\ldots,m-1, \beta = 0,\ldots,k-1, \gamma = 0,\ldots,n-1$ and $p$ denotes the number of computing nodes. 

However, in some cases, where $p \gg m$, sharding w.r.t $\alpha$ may cause severe computational unsaturation. An improved form of $h_{rand}$ can be introduced as:
\begin{displaymath} h_{rand}(\alpha,\beta,\gamma) = hash(\alpha,\beta,\gamma) \mod p\end{displaymath}
where the function $hash$ calculates hash of all three parameters, and guarantees uniformity for its result, such that shards and computation can be equally distributed to each machine. Unfortunately, using this function, data locality is violated, since this sharding policy depends on all its three parameters rather than $\alpha$ itself, thus, a secondary shuffling may be triggered before the summation stage. 

\subsubsection{Partitioning Strategy}
\label{sec:partstr}
The sparsity maybe the first factor that should be taken into account since the sparsity of matrices output may largely affect the efficiency of the ``Shuffle" stage where the intermediate results are combined and aggregated. As noted before, for partition schema $m,n,k$ partitioning duplicates original matrices in $m \times k$ times, and the size of intermediate results before final summation is $n$ times of final output so that for a large and dense matrix, the size of both partition groups and intermediate results may be too large to be transmitted via network. 

The second concern is the profile (e.g. width and height) of matrices being multiplied. Generally, the number of partition should be as small as possible so that not too many partition groups are generated and small amount of intermediate results are emitted, however, the number of partition should be large enough so that computational tasks can be distributed equally, and parallel computing can be fully utilized.

\begin{algorithm}
\caption{Summation Process (Mapper)}
\label{alg2}
\begin{algorithmic}
\REQUIRE partition group identifier $\langle\alpha, \beta, \gamma, i \rangle $
\REQUIRE aggregated sub matrices $\{\mathbf{A}_{sub}, \mathbf{B}_{sub}\}$

\FOR{\textbf{each row} $\mathbf{c} \in \mathbf{A}_{sub}\mathbf{B}_{sub}$} 
\STATE $emit(i, \mathbf{c})$
\ENDFOR
\end{algorithmic}
\end{algorithm}

\begin{algorithm}
\caption{Summation Process (Reducer)}
\label{alg3}
\begin{algorithmic}
\REQUIRE row index $i$, partial results $\mathbf{C} \in \{c_{i1},c_{i2},\ldots,c_{in}\}$ \PF{should this be $\mathbf{C} = \{c_{i1},c_{i2},\ldots,c_{in}\}$?} \SL{yes}
\STATE $\mathbf{c_i} \gets 0$ \PF{shouldn't this be 0?} \SL{yes}
\FOR{\textbf{each} $\mathbf{c \in C}$} 
\STATE $\mathbf{c_i = c_i+ c}$
\ENDFOR
\STATE $emit(i, \mathbf{c_i})$

\end{algorithmic}
\end{algorithm}

\subsection{Iterative Multiplicative Update}
Compared with the earlier approach, the implementation for this update is much simpler and straightforward. Since the matrix $\mathbf{B}$ is often small enough to fit into memory and is very convenient to be transmitted among machines. In this implementation, they are distributed to each machine before each iteration, so that during the computation process, no communication is required. Therefore, the matrix $\mathbf{A}$ can be split into random pieces as long as each entire row is preserved. For each row $\mathbf{r_i}$ in segment of matrix $\mathbf{A}$, row $\mathbf{c_i}$ in final result can be obtained by 
\begin{equation}
\label{eq:iterativemul}
\mathbf{c_i} = \mathbf{r_iB}
\end{equation}
The mechanism of ``Distributed Cache" provided by Hadoop can be regarded as a possible approach for distributing small matrix $\mathbf{B}$ among computing nodes. This functionality allows servers only load matrix once when initiating the calculation, and no communication is required afterwards.

\subsection{Algorithm-Specific Settings}
Although two general models have been extracted from three algorithms, for each algorithm, specific settings still need to be discussed for performance optimization. 

\subsubsection{Non-Negative Matrix Factorization}
Shown by Definition~\ref{def:nmf}, the essence of NMF is to decompose one large matrix as a multiplication of one ``tall" matrix ($m \gg k $) and one ``fat" matrix ($k \ll n$), according to ~\eqref{eq:nmflinear}, solution of NMF involves both types of multiplicative models. 
It follows that $\mathbf{A H^{T}}$, $\mathbf{H H^T}$, $\mathbf{W^T A}$ and $\mathbf{W^T W}$ can be classified as ``Similarity Measure Multiplication" as mentioned above and considering the size and sparseness of different operands matrices, optimization can be made when the unique profiles of these matrices are utilized. 
Since these four multiplications are symmetric, for simplicity, only $\mathbf{W^{T} A}$ and $\mathbf{W^T W}$ are discussed in this section. 

Typically, $\mathbf{W^{T} A}$ handles one large sparse matrix $\mathbf{A} \in \mathbb{R}^{m \times n}$ and a fat matrix $\mathbf{W^T} \in \mathbb{R}^{k \times m}$, and generates a dense fat matrix where $k \ll m,n$. A sensible partition schema should avoid splitting across columns(the shorter edge) of $\mathbf{W^T}$. In contrast partition on rows(the longer edge) is a good plan, since a entire row may be simply too large to be fit to memory. Besides, splitting across columns may only have little effects on performance but largely increases the partitioning workload. 

Similarly, the partition schema of $\mathbf{W^T W}$ can also be formed when considering factors listed above. Matrix $\mathbf{W} \in \mathbb{R}^{m \times k}$ and its transposition which are ``tall" and ``fat" matrices respectively, multiply and generate a very small matrix $\mathbf{C} \in \mathbb{R}^{k \times k}$. For the reason stated above, splitting across rows of $\mathbf{W^T}$ is preferred, while each column vector of $\mathbf{W^T}$ or row vector of $\mathbf{W}$ should be remained as a unity.

In order to compute $\mathbf{Y=W^TWH}$, our second multiplicative model can be adopted when the small matrix $\mathbf{C}$ has been obtained from $\mathbf{W^T W}$. In normal cases, it is feasible to distribute $\mathbf{C}$ among machines. Several minor changes have to be conducted before this calculation can be fit into \eqref{eq:iterativemul}. To compute each row $\mathbf{Y_i \in Y}$, entire row $\mathbf{C_i \in C}$ and column $\mathbf{H_j \in H}$ must be accessible for the machine where $\mathbf{Y_i}$ is going to be calculated. However, traditional row-based file format only allows us to retrieve matrix by row. This problem can be solved by transposing matrix $\mathbf{H}$, and each column $\mathbf{Y_j \in Y}$ can be obtained by computing $\mathbf{Y_j=H_jC^T}$, where $\mathbf{H_j}$ is read from transposed matrix $\mathbf{H^T}$. In fact, since each single column $\mathbf{Y_j}$ is sequentially stored in final output, we actually compute the $\mathbf{Y^T}$ instead of $\mathbf{Y}$. 


\subsubsection{Support Vector Machines}
Kernel computation is a typical pairwise similarity comparison that stores results in the form of kernel matrix which is a large dense matrix grows quadratically with size of training set. Training set $\mathbf{T} \in \mathbb{R} ^{a \times b}$ is formed as a matrix where each training example $\mathbf{T_i}$ is represented as a vector with $b$ dimensions. Linear kernel $\mathbf{K} \in \mathbb{R} ^{a \times a}$ can be calculated as 
\begin{displaymath}\mathbf{K=TT^T}\end{displaymath}
where $\mathbf{T}$ is usually a sparse matrix. 

Since both $a$ and $b$ may be large in practical, partition schema may take both dimensions into account. The real challenge is the large data storage before the final result is summed. Because of the density of matrix $\mathbf{K}$, the size of intermediate results may grow linearly \PF{find a better word here!}\SL{fixed}when the number of partitions on $b$ grows, however, although partition on $a$ may also duplicate partitioned groups, matrix $\mathbf{T}$ is usually very sparse so that duplication may have less space requirement. In conclusion, a larger $m,k$ is preferred to a larger $n$ in this case. 

The SVM training process can be conducted by iterative multiplicative process (see  \eqref{eq:linearsvm}). 

\subsubsection{PageRank}
As mentioned above, the power method \eqref{eq:pageranklinear} can be expressed very well using iterative multiplicative model, no specific engineering on our framework is needed. 
However, there may be some issues when the ``Damping Factor" is introduced. Limited by  space, no further related discussion will be made in this paper. 

\section{Experiments and Results}
Experiments in this research are conducted in two phases: \textit{a)} experiments of similarity multiplication, \textit{b)} experiments of parallel algorithms. 

In the first phase of our experiments, we focus on matrix multiplication in
order to explore the potential optimal setting among a wide range of parameters and reveal the
relationship between time performance and the size, sparsity, and partitioning strategy of
input matrices. The final objective is to form a practical guide for choosing optimal
parameters.

The next stage of the experiment is designed for the adapted algorithms. These experiments
are conducted for two purposes: a) verifying the correctness of our implementations, b)
illustrating performance impact brought by parallelization. Evaluations and analyses will be
given after each group of experiments respectively.

\subsection{Experimental Environment}
As a mature MapReduce implementation, Hadoop\footnote{http://hadoop.apache.org/} is selected as our experimentation platform, and algorithms are implemented in Java under JDK 1.6 runtime. 
All experiments mentioned in this section are performed on the BlueCrystal High
Performance Computing Cluster\footnote{https://www.acrc.bris.ac.uk/}. Having achieved a performance of 28.4 TFlop/s, BlueCrystal was placed 86th in the Top500 in June, 2008.\footnote{https://www.acrc.bris.ac.uk/acrc/hpc.htm}

\subsection{Matrix Multiplication Experiments}

\subsubsection{Dataset Construction}
With the help of MapReduce, a large matrix $A \in \mathbb{R}^{m \times n}$ can be generated in parallel using the matrix generator with the MapReduce algorithm given in Algorithm~\ref{alg4} and Algorithm~\ref{alg5}, where $\delta$ is the sparsity defined by the fraction of non-zero elements in all elements. By default, $m=n=2^{13}, \delta=2^{-7}$.

\begin{algorithm}
\caption{Matrix Generator (Mapper)}
\label{alg4}
\begin{algorithmic}
\REQUIRE matrix height $m$
\FOR{$i \gets 1$ \TO $m$} 
	\STATE $emit(i, \{\} )$
\ENDFOR
\end{algorithmic}
\end{algorithm}

\begin{algorithm}
\caption{Matrix Generator (Reducer)}
\label{alg5}
\begin{algorithmic}
\REQUIRE row index $i$, sparseness $\delta$, matrix width $n$
\STATE $\mathbf{row} \gets \{\}$
\FOR{$1$ \TO $n$} 
\IF{$random() < \delta$}
\STATE $\mathbf{row} \gets \mathbf{row} \cup \{ random() \} $
\ENDIF
\ENDFOR
\STATE $emit(i, \mathbf{row} )$

\end{algorithmic}
\end{algorithm}

In this group of experiments, the height and width are equal, and both are in a factor of 2 in order to demonstrate the potential non-linear relationship between experiment variables. The square matrix is also straightforward in the representation of space and time complexity. 

The major advantage of generating a matrix by using MapReduce is the output matrix from a generator will be well balanced among all the computing nodes, which means the work load for each Mapper will not be significantly distinguished. Therefore, the data locality can be enabled at the beginning.

This generator is also employed for other purposes to generate particular matrix (e.g. random initial values), which may have different requirements compared with the experiments listed here. Details are not listed in this paper. 

\subsubsection{Computational Complexity w.r.t $m$}
\label{sec:complexity}
Theoretically, for multiplication of two square matrices $\mathbf{A,B} \in \mathbb{R}^{m \times m}$, the computational complexity is $O(m^3)$ that dominates among all matrix operators. Implemented from the naive algorithm, the complexity of our operator remains the same. However, in most cases (particularly in Text Mining applications), the matrices are often extremely sparse although they usually have very high dimensions. Considering this fact, the actual computational complexity should be much reduced than the theoretical complexity when sparse matrix algorithm is used. 

In this experiment, the performance of a matrix multiplication is investigated through 6 tests, each of which performs a multiplication between two randomly generated squared matrices $ \mathbf{A,B} = \mathbb{R}^{m \times m}$. The parameters remain at their default settings.

\begin{figure}
\centering
\includegraphics[scale=.4]{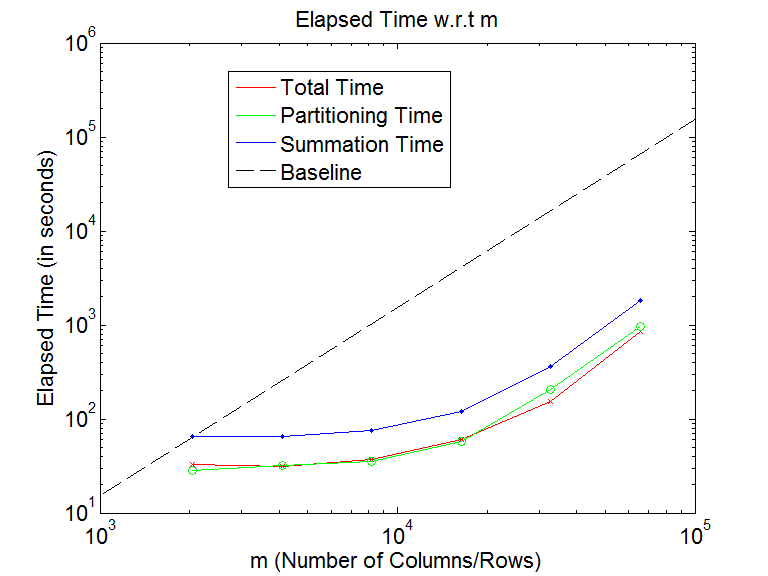}
\caption{Elapsed time wrt.\ the number of rows/columns plotted in Log-Log space. In this picture, baseline are dashed  lines through the first blue points with slopes = 3. Partitioning time and summation time are also drawn along with the total elapsed time.}
\label{fig:matmulti1}
\end{figure}
\PF{this bit needs improving}
Figure 2 generally generally confirms the assumptions we have made earlier. Three lines are plotted on the graph which show the increasing trend of running time, partition time and summation time when the size of matrix increases. Beyond $m=131,072$ we reach the storage limit of the HPC cluster. As can be expected from the $O(m^{3})$ complexity of matrix multiplication, all lines grow non-linearly with increasing slopes. The ``Baseline'' is plotted as a theoretical benchmark which illustrates the theoretical prediction of elapsed times for the each experiment. 


From Figure~\ref{fig:matmulti1} we can see that the gap between the theoretical prediction and actual results are large. For all sections of the curves, slopes are lower than predicted. Even at the final stage of the curves, the slope is approximately 2 compared with 3 of the prediction baseline. This gap can be explained by the sparsity of input matrices for which a large number of zero cells are not stored and actually computed. 

Figure~\ref{fig:matmulti1} also illustrates that the total running time is equally shared between the partitioning and calculation stage.

\subsubsection{Performance w.r.t $\delta$}
\label{sec:sparsity}
\begin{figure}
\centering
\includegraphics[scale=.4]{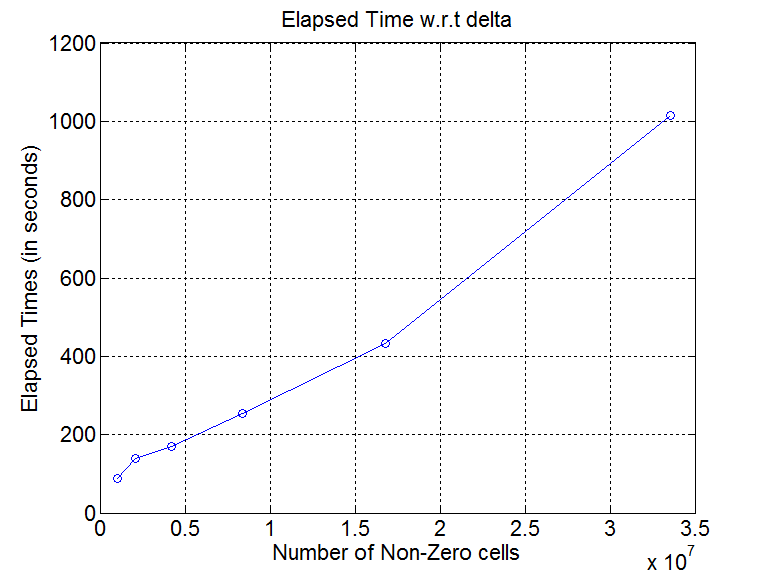}
\caption{There is an almost linear correlation between elapsed time and the number of non-zero cells in operand matrices.}
\label{fig:matmulti2}
\end{figure}

In order to discover the correlation between the elapsed time and the number of non-zero cells in $\mathbf{A, B}$, we plot the $T_{elapse}$ vs sparsity in the Figure~\ref{fig:matmulti2}. The whole experiment is performed on two randomly generated square matrices where $ m = 2^{14}$. \PF{should probably be $2^{-12}$} \SL{fixed}

From this picture it can be seen that the elapsed time grows linearly when the number of non-zero cells increases.

The linear pattern can also be explained when looking back at Figure~\ref{fig:matmulti1}. Since the elapsed time and the number of non-zero cells are both quadratically correlated with $m$, a linear relationship can be expected between these two variables.

\subsubsection{Performance w.r.t Partitioning Function and Schema}
As has been mentioned earlier, choosing different partitioning strategies can have significant impact on performance. The naive partitioner is developed for maximizing the usage of data locality. 
This experiment demonstrates how much benefit is obtained by the naive partitioning strategy. We use the same settings employed in section~\ref{sec:complexity}. However, the experiments in this section are repeated three times with different partitioning functions: \textit{a)} $h_{rand}$ with partition schema $\langle m=20, n=6, k=20 \rangle $, \textit{b)} $h_{rand}$ with partition schema $\langle m=40, n=6, k=40 \rangle $ , \textit{c)} $h_{naive}$ with partition schema $\langle m=20, n=6, k=20 \rangle $.

\begin{figure}
\centering
\includegraphics[scale=.33]{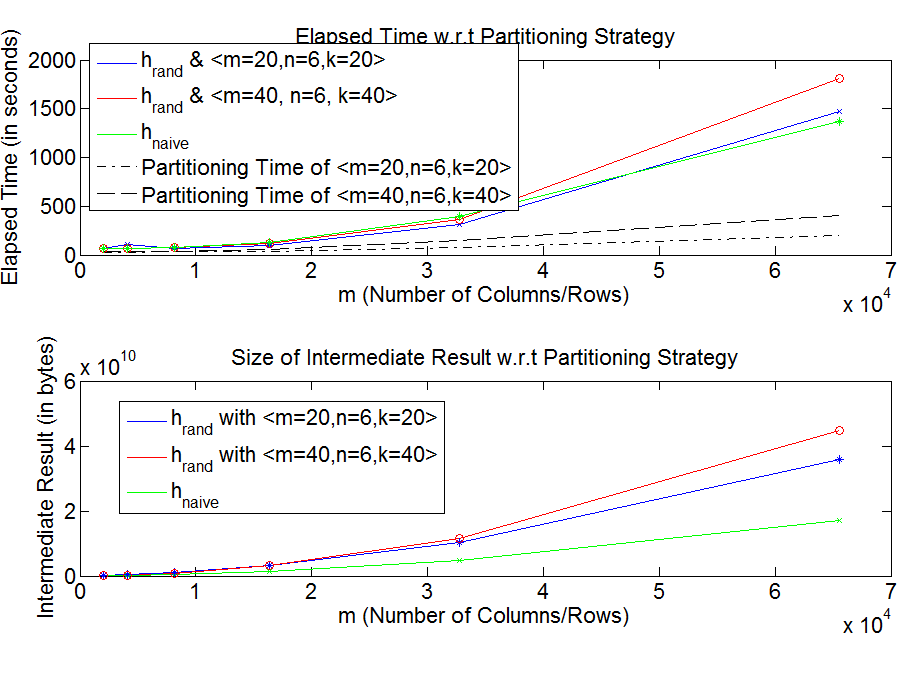}
\caption{Space and time performance of three different partition strategies used in matrix multiplication. Partitioning time for two different random partitioners also illustrates on the first graph.}
\label{fig:matmulti3}
\end{figure}

From the two charts in Figure \ref{fig:matmulti3}, we can see clearly that compared with the other two partitioning strategies, $h_{naive}$ enjoyed less intermediate results and smaller $T_{elapse}$ on both pictures, as we have predicted in section \ref{sec:partstr}. The $h_{naive}$ has a great potential in reducing the size of intermediate results for the usage of data locality. It is also not surprising that $h_{rand}$ with larger $\langle m, n, k \rangle $ leads to more intermediate results for the reasons stated in section \ref{sec:partsch}.

\subsubsection{Speedup Rate}
\begin{figure}
\centering
\includegraphics[scale=.4]{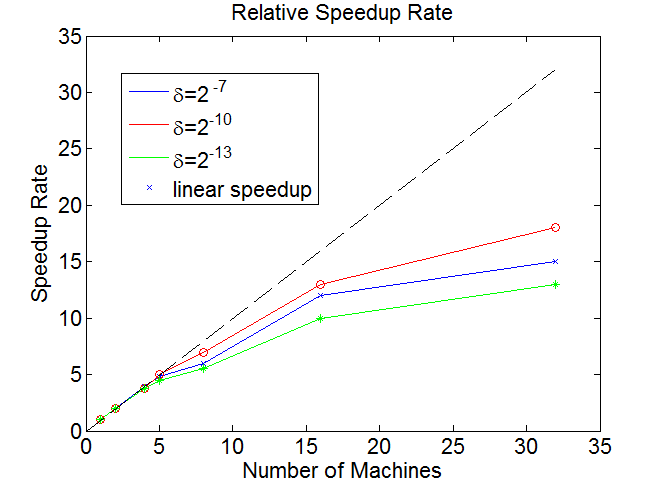}
\caption{Speedup rates of three different $\delta$ choices. The ideal speedup is illustrated by the diagonal line.}
\label{fig:matmulti4}
\end{figure}

Considering the capacity of a single machine and the software available, only the relative speedup is adopted for the test in this experiment.
From the graph in Figure \ref{fig:matmulti4}, a clear trend of speedup rate can be seen. The speedups with three different $\delta$ are all lower than the linear speedup which upper-bounds all the practical speedup according to {Amdahl's law}.
On the matrix with $ \delta=2^{-7}$ , the linear speedup is almost 7 when 8 workers are enabled. The other curves with $\delta= 2^{-10}$ and $\delta= 2^{-13}$ also have similar speedup rates when the number of working machines is less than 8.

The reasons for the gap between the practical speedup and the theoretical upper bound can be various. Although the code we wrote in MapReduce fashion can be all parallelized, several maintenance operations may be conducted during the experiment, such as check-pointing and auto-balancing.
Interestingly, we can observe that speedup for sparser matrices are lower than those with high density.\PF{This doesn't make sense: sparser is the same as higher sparsity!}\SL{Right, I mean high density}
This fact suggests that for small and sparse matrices, a small number of clusters can do the best job, while for larger clusters, the whole system is not saturated and the rest of the computing resources (e.g. memory and idle CPUs) are wasted.
\subsection{Experiments of Parallel Algorithms}
In this section, the correctness and parallel performance of our algorithm implementations are tested against a range of datasets.

\subsubsection{Datasets}
In considering two different experiment objectives, two groups of problem sets are chosen for different purposes.
For correctness experiments, we test our implementation against a series of small problem sets which are fit for both our parallel implementation and sequential algorithms.
For parallel experiments, different test sets are selected for different problems settings of algorithms. During the SVM experiments, a typical text classification problem is tested by using the Reuters Corpus\footnote{http://about.reuters.com/researchandstandards/corpus/}  with various number of training examples ranging from 12k to 192k, while for NMF, a random generated matrix is used for factorization process, and finally, Wikipedia 2008 corpus\footnote{http://www-connex.lip6.fr/~denoyer/wikipediaXML/}  which contains 5,716,808 compressed wiki-pages from Wikipedia English Website are used for PageRank link analysis. 

\subsubsection{Experiments on SVM}
We use LibSVM \cite{CC01a}\PF{need citation} \SL{fixed} for comparison because it is widely used and has been tested against many publicly available datasets. In this experiment, it is set to the default model without using any optimization techniques.
Table~\ref{table:svm1} shows accuracy comparison of LibSVM and the Multiplicative Models. Linearly separable datasets are marked by *.
As can be seen from the table shown in \ref{table:svm1}, for two linear separable problem sets, our implementation offers very close accuracies compared with those achieved by LibSVM. The best performance is reached on the IRIS dataset with 98.7\% compared with 97.33\% from LibSVM. Besides, for non-linear separable problems (Reuters-21578 and ADULT), our implementation still achieves a promising accuracy (69.3\% and 87.2\% respectively), although LibSVM enjoys a much higher accuracy for its soft-margin classification implementation.

For parallel experiments, our SVM trainer will be trained on a series of subsets of Reuters Corpus for a fixed iteration number\footnote{Stopping criteria for SVM are beyond the scope of the current paper.}  with increasing number of instances. The experiments start with 12K training examples, and end up with 192K training examples. The size of kernel matrix generated and training time for one iteration (one multiplicative process) is reported in table~\ref{table:svm2}. 

The first two columns listed in table~\ref{table:svm2} illustrate the major issue that SVM suffers from. Dense and large kernel matrix makes the SVM less attractive for large-scale problem sets. All kernel matrices that listed above have a very high density \PF{do you mean density?} \SL{yes} level (above 0.8). Moreover, the size of the kernel matrix also grows with the number of training examples quadratically. For sequential computing and traditional gradient descent, these kernel matrices are not acceptable for in-memory access. 

The last two columns report the elapsed time for each iteration and accuracy. Generally, our algorithm can obtain an acceptable accuracy when adopted on large scale of dataset within a reasonable time. Interesting patterns can be found in the last column where the accuracy is dropping as the number of instances increase. We assume this may be caused by our stopping criterion that terminates our algorithm at an immature stage, however, we can not find a very strong proof for our guess, and we believe this remains a tasks for future study.

\begin{table}
\caption{Accuracy comparison with sequential Libsvm}
\label{table:svm1}
\centering
\begin{tabular}{|c|c|c|}
\hline
Dataset&Libsvm &Multiplicative Models\\
\hline\hline
IRIS* &97.33\% &98.7\%\\
ADULT &82.1\% &69.3\%\\
WINE* &98.5\% &95.2\%\\
HEART (Binary) &97.6\% &95.2\%\\
Reuters-21578 &92.2\% &87.2\%\\
\hline
\end{tabular}
\end{table}

\begin{table}
\caption{Parallel performance of SVM on Reuters Corpus}
\label{table:svm2}
\centering
\begin{tabular}{|p{1.4cm}|p{1.6cm}|p{1.3cm}|p{.6cm}|p{1.3cm}|}
\hline
NO. Training Examples
&Kernel sparsity
&Kernel Matrix
&$T'$
&Accuracy\\

\hline\hline
12K &0.82 &1.47 &33s &82.5\%\\
24K &0.83 &5.70 &45s &80.3\%\\
48K &0.86 &22.3 &66s &74.2\%\\
96K &0.87 &74.0 &180s &72.4\%\\
192K &0.86 &230 &420s &70.0\%\\
\hline

\end{tabular}
\end{table}

\subsubsection{Experiments on NMF}
The aim of this set of experiments is to show our implementation has gained the full capability of solving the large-scale matrix-factorization problem on HPC cluster. Since the update of W and H are symmetric, in this section, only performance of updating H will be discussed. 

The performance is measured by three different computational components in our algorithm. For each component, four conditions under two categories will be considered. Similar to the evaluation of matrix multiplication, the effect of matrix sparsity is of interests. Moreover, the parameter $k$ will also be considered under each sparsity setting.

The first interesting pattern that can be found among the data listed in the table~\ref{table:nmf1} is the elapsed time of computing $\mathbf{X=W^T A}$ dominates both computational costs in terms of both sparsity settings. The reason for this is that the matrix A is commonly larger than both $\mathbf{W}$ and $\mathbf{H}$, and the partitioning may have a higher cost compared with others. For this reason, we can expect a drop of elapsed time when $\mathbf{A}$ becomes sparser. This expectation has been verified by the other half of our table, which shows, $T_{elapse}$ reduces dramatically when the
sparsity decreases to $2^{-10}$ . The performance of other components drops accordingly. The general trend in this table fits well with the results reported in \cite{liu10}.

However, it can also be observed that the computation of $\mathbf{H=H.*X./Y}$ is almost a constant, although the computational complexity of this component should be $O(n)$. It suggests that the capacity of the current cluster is not saturated. In this case, it may be more efficient to run on local machines than in a distributed environment.

\begin{table}
\caption{Parallel performance of NMF on Random Matrices}
\label{table:nmf1}
\centering
\begin{tabular}{|p{2cm}|c c | c c|}
  \hline
  \multirow{3}{2cm}{Computational Components}& 
  \multicolumn{2}{c|}{$\delta = 2^{-7}$}& 
  \multicolumn{2}{c|}{$\delta = 2^{-10}$}\\ 
  
  &  $k=8$ &  $k=32$ &  $k=8$ &  $k=32$ \\
  &  $T_{elapse}$ &  $T_{elapse}$  &  $T_{elapse}$  &  $T_{elapse}$ \\
  \hline \hline
  $\mathbf{X=W^T A}$ & 46 & 129 & 20 & 30\\
  $\mathbf{Y=W^T W H}$ & 23 &60 &13 &16\\
  $\mathbf{H=H.*X./Y}$ & 12 &14 &11 &12\\

  \hline
\end{tabular}
\end{table}

\subsubsection{Experiments on PageRank}

We first test our implementation on two small datasets: \textit{a)} Harvard500, which is a directed graph adjacent matrix generated from 500 webpages crawled from Harvard University website. \textit{b)} Hollions.edu, which is also a web matrix organised from crawled web-content on Hollins.edu. As mentioned before, we use the Wikepedia 2008 Corpus as our parallel testing dataset. 

\begin{figure}
\centering
\includegraphics[scale=.4]{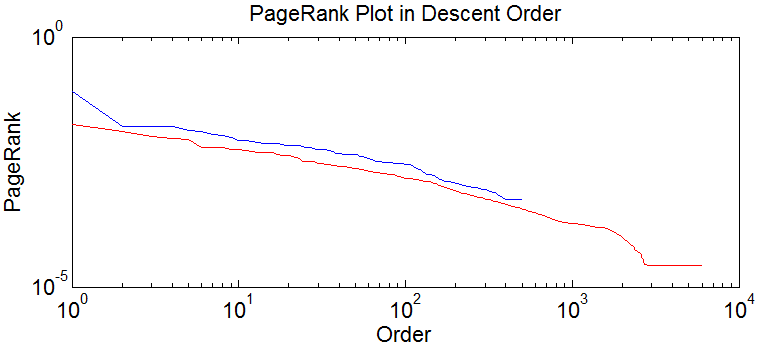}
\caption{PageRank plot of Harvard500 and Hollins.edu respectively in descending order}
\label{fig:pagerank}
\end{figure}

After adopting our PageRank calculation on both datasets, the obtained          PageRank array is sorted in descent order, and plotted into Log-Log space in Figure~\ref{fig:pagerank}.

It can be noticed that the PageRank of Hollins.edu is smaller than Harvard500, since the number of webpages (i.e. row/column in web-matrix) in Harvard500 is much smaller than that in Hollins.edu. (According to the algorithm, the sum of all PageRank should be exactly 1. )

\begin{figure*}
\centering
\includegraphics[scale=.37]{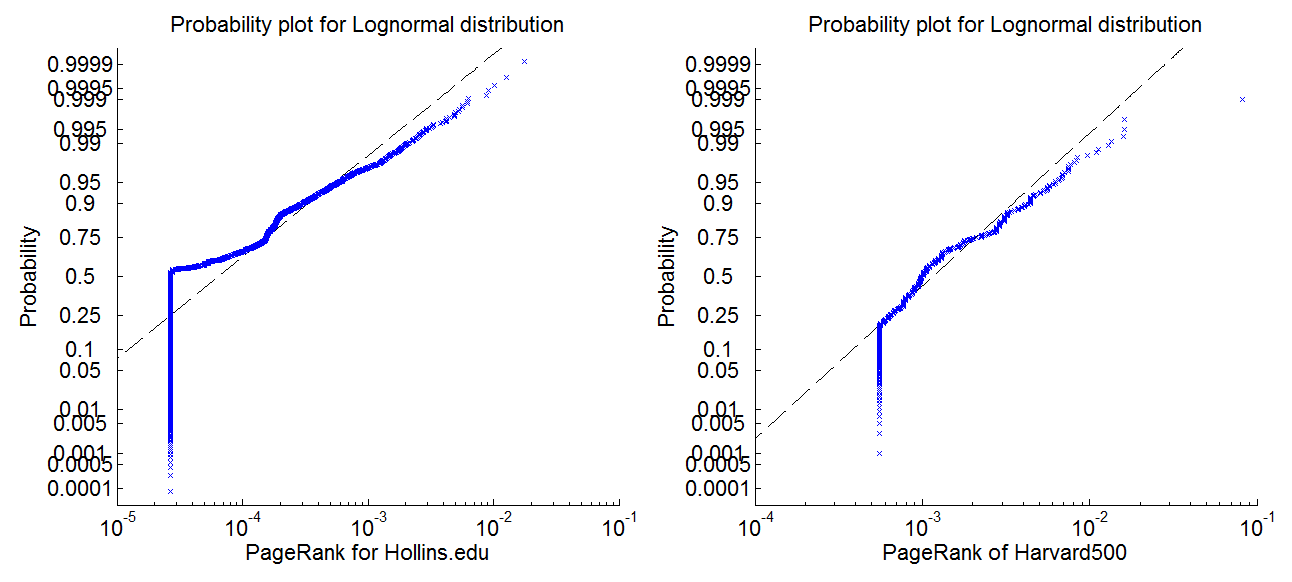}
\caption{Probability plot of PageRank for lognormal distribution}
\label{fig:pagerank2}
\end{figure*}

The PageRank for both datasets reduces linearly in LogLog space. It suggests that the PageRank of these two datasets conforms to certain logarithm distribution, and this assumption can be verified by two probability plots shown in Figure~\ref{fig:pagerank2}.

From the two charts in Figure~\ref{fig:pagerank2}, it can be seen that PageRank on both datasets fits the lognormal distribution well except for samples on the left hand side, where the PageRank is lower than a certain value and remains equal to each other. This sign may suggest that the PageRank on both datasets has not fully converged for small values. 

\begin{figure*}
\centering
\includegraphics[scale=.27]{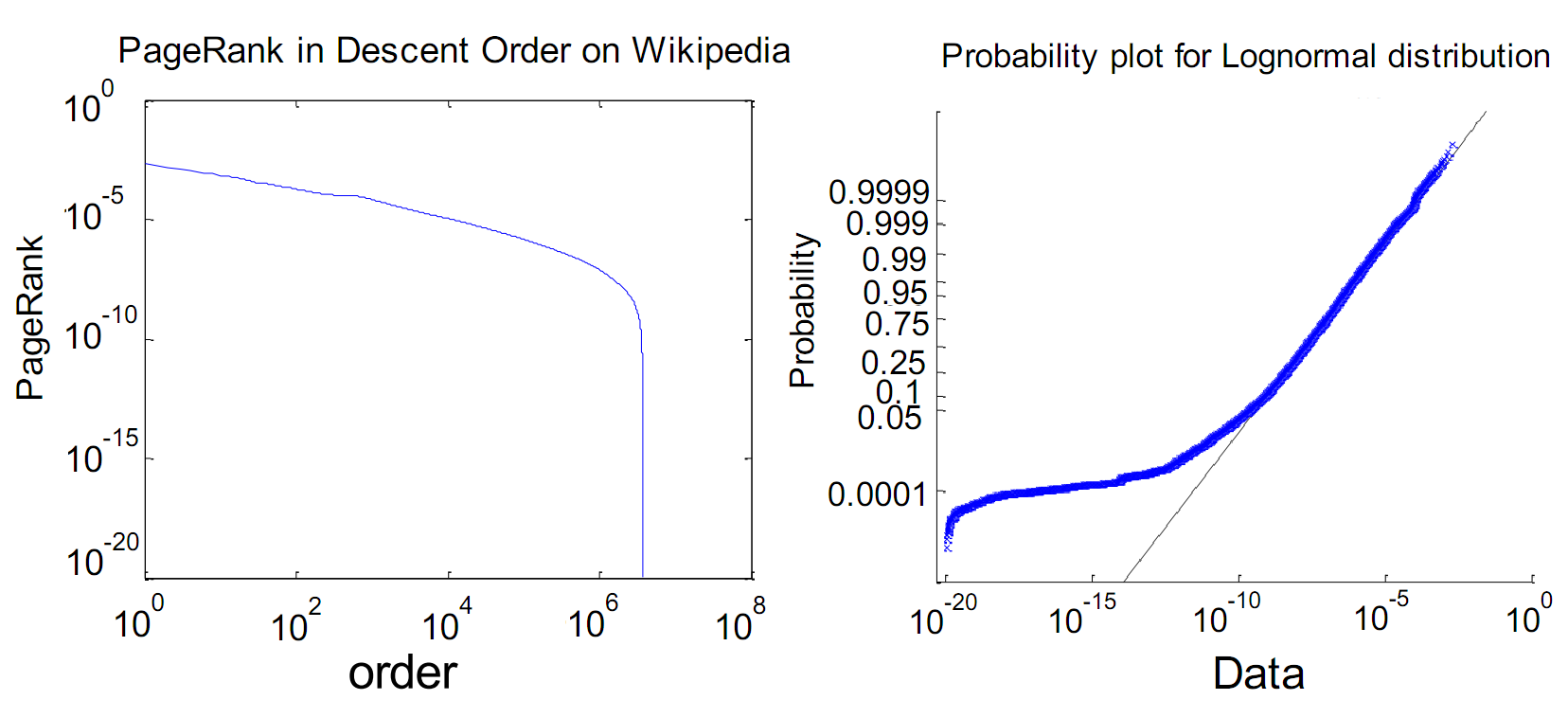}
\caption{PageRank plot and probability plot on Wikipedia Corpus}
\label{fig:pagerank3}
\end{figure*}

In order to evaluate performance on real-world dataset, our algorithm is executed on the Wikipedia dataset in parallel. By using 30 nodes, our implementation finishes 100 iterations within 1 hour, and provides us with the following results shown in Figure~\ref{fig:pagerank3}.

The two graphs plotted in Figure~\ref{fig:pagerank3} show different patterns compared with those generated on small datasets. In Log-Log space, the PageRank falls suddenly after the $10^{6th}$, which implies that there is a small collection (approx. $2 \sim 3 \times 10^6$ ) of Wikipedia entries not connected with the
most regular entries below a certain level of PageRank (approx. $10^{-7}$). The probability
distribution shows that top ranks also conform to Lognormal Distribution, however, it can be noticed that there is a large PageRank shift for the smallest ranked group of entries. It can be regarded as another interpretation of the pattern on the left chart in Log-Log space.

Another interesting result is reported in the table~\ref{table:pagerank1} which lists the top 10 results sorted by PageRank.

Although Wikipedia does not release its official popularity ranking, measured by our daily
life experience, a reasonable popularity list seems to be produced by our PageRank implementation. However, it also gives an arguable rank of entry \textit{Geographic Coordinate System}.

Generally, as the PageRank of webpage decreases, its number of inlinks also drops. The only
exception is the entry \textit{Wiktionary}, which is ranked 5th, while its inlinks number fails to list in the Top 100 entries sorted by No. inlinks, thus it is marked by ``\textbf{$<29076$}" (where 29076 is the smallest inlink number among Top 100).\\ \\

\begin{table}
\centering
\caption{Top-10 ranked Wikipedia Entries with their PageRank and NO. inlinks}
\label{table:pagerank1}
\begin{tabular}{|c|c|c|}
\hline
Wikipedia Entry&PageRank &NO. Inlinks\\

\hline\hline
United States & 0.0021 & 374934\\
2007 & 0.0014 & 266614\\
2008 & 0.0014 & 286409\\
United Kingdom & 0.0011 & 139325\\
Wiktionary & 0.0009 & < 29076\\
Geographic coordinate system & 0.0009 & 294604\\
2006 & 0.0009 & 146336\\
Wikimedia Commons & 0.0008 & 70096\\
English language & 0.0006 & 69408\\
Germany & 0.0006 & 95366\\
\hline

\end{tabular}
\end{table}

\section{Conclusion and Future Works}
In this paper, three machine learning algorithms: Non-Negative Matrix Factorization, Support Vector Machines, and PageRank have been successfully adapted to the new parallel paradigm MapReduce for large-scale problems. 

To achieve this goal, two generic multiplicative components are extracted from the solutions to these problems. We further discussed the unique features of these models concerning different parallel settings. A general configuration of ``Partition Schema" has been introduced to describe the partition strategies under different sparsity/density settings and various scales of clusters.

Our main contribution in this study is to propose a generic model that is efficient for upscaling a set of learning algorithms, particularly those involving distances and similarity measures, and iterative multiplicative updates. Compared with upscaling different algorithms individually, we believe our approach provides one general solution to the large-scale learning problems.

Analysis from our experiments shows that all our models are correctly adapted and successfully applied to three algorithms. For partitioned matrix multiplication, we observed a encouraging linear speedup and reduced computational time compared with theoretical prediction. An optimization brought by partitioning schema has also been discovered. Our experiments also show that the density of matrices being multiplied is a crucial factor for speedup rate.

Three learning algorithms have also been tested for correctness and performance. Results show NMF also achieves a good speedup rate, especially for the portion of computation related to similarity comparison multiplication. SVM achieves a comparable performance to the mainstream SVM toolkit LibSVM while also revealing good scalability on extremely large kernel matrices. The PageRank implementation was successfully applied on the Wikipedia corpus with 5,716,808 entries for 100 iterations within 1 hour, and shows encouraging results which are ``empirically" good. 

Several lines of future work suggest themselves. Limited by the simple functional representation of MapReduce, we cannot at the moment implement more sophisticated algorithms which may involve instant machine communications. Simply, the ``online version" of MapReduce may be worth considering for the future development on MapReduce structure. Once the ``online" version has been implemented, a large collection of stream-based algorithms especially involving real-time updating techniques may become possible targets on MapReduce. Fortunately, a prototype ``MapReduce Online" has been designed in \cite{condie10}, and a Hadoop version\footnote{http://code.google.com/p/hop/} is also available. 

\PF{need a better concluding sentence} \SL{look better?}
To conclude, both theoretical and experimental results show that our generic approach is widely applicable for upscaling similarity and iterative-based learning problems on MapReduce and reveals potential on data sets with various scales. Performance may be further improved by setting up a fitting partition schema. 

\section{Open Source Implementation}
The techniques discussed in this paper forms the prototype of project BigO2 (http://code.google.com/p/bigo2/).

\section{Acknowledgments}
This work was carried out using the computational facilities of the Advanced Computing Research Centre, University of Bristol - http://www.bris.ac.uk/acrc/.

\bibliographystyle{plain}
\bibliography{ref}




%

\end{document}